\documentclass[useAMS,usenatbib,usegraphicx,onecolumn]{mn2e}
\usepackage{graphicx}
\usepackage{amsmath, amssymb}

\makeatother
 
\begin{document}

\title[The onset of warps]{The onset of warps in {\em Spitzer} observations of edge-on spiral galaxies\\}

\author[Saha et al.]{Kanak Saha$^{(1,2)}$\thanks{E-mail:kanak@physics.iisc.ernet.in}, Roelof de Jong$^3$ \& Benne Holwerda$^4$\\
$^{1}$Department of Physics, Indian Institute of Science, Bangalore 560012, India\\ $^{2}$Academia Sinica Institute of Astronomy and Astrophysics - TIARA, P.O. Box 23-141, Taipei 10617, Taiwan\\ $^{3}$Space Telescope Science Institute, Baltimore, MD21218, USA\\$^{4}$ Department of Astronomy, University of Cape Town, Republic of South Africa }

\maketitle

\begin{abstract}

We analyze warps in the nearby edge-on spiral galaxies observed in the {\em Spitzer/IRAC} 4.5 micron band. In our sample of 24 galaxies we find evidence of warp in 14 galaxies. We estimate the observed onset radii for the warps in a subsample of 10 galaxies. The dark matter distribution in each of these galaxies are calculated using the mass distribution derived from the observed light distribution and the observed rotation curves. The theoretical predictions of the onset radii for the warps are then derived by applying a self-consistent linear response theory to the obtained mass models for 6 galaxies with rotation curves in the literature. By comparing the observed onset radii to the theoretical ones, we find that discs with constant thickness can not explain the observations; moderately flaring discs are needed. The required flaring is consistent with the observations. Our analysis shows that the onset of warp is not symmetric in our sample of galaxies. We define a new quantity called the onset-asymmetry index and study its dependence on galaxy properties. The onset asymmetries in warps tend to be larger in galaxies with smaller disc scale lengths. We also define and quantify the global asymmetry in the stellar light distribution, that we call the edge-on asymmetry in edge-on galaxies. It is shown that in most cases the onset asymmetry in warp is actually anti-correlated with the measured edge-on asymmetry in our sample of edge-on galaxies and this could plausibly indicate that the surrounding dark matter distribution is asymmetric.  
\end{abstract}


\section{Introduction}
The warps in spiral galaxies continue to be a puzzling phenomenon even after about half a century since their discovery (Kerr 1957). Recent observational analysis along with theoretical studies have substantially improved our understanding of the nature of warps in spiral galaxies. Warps are primarily seen in the 21 cm neutral hydrogen line (Bosma 1978; Briggs 1990; Garcia-Ruiz et al. 2002) and the optical bands (Sanchez-Saavedra et al. 1990; Reshetnikov \& Combes 1999; Drimmel et al. 2000; Reshetnikov et al. 2002). Based on the IRAS (Infrared Astronomical Spectrograph) observations of the old stellar population, Djorgovski and Sosin (1989) detected warps in the stellar disc of our Galaxy and the signature of warps in the mid-infrared has also been confirmed recently by Vig et al. (2005). However, unlike the optical, the infrared (IR) observations and a thorough analysis of warps in external galaxies are lacking, a situation we try to remedy here. 

The amplitudes of warps in the optical and infrared are much smaller than in the H{\footnotesize I}. Based on a small sample of HI observations of warps, Briggs (1990) created a set of rules for warps. His first rule states that warps develop between R$_{25}$ and the Holmberg radius, R$_{Ho}$ (corresponding to 25 and 26.5 B-mag arcsec$^{-2}$ surface brightness respectively). Typically, these radii would correspond to 4 - 6 exponential radial scale lengths. However, no such rule exists so far in the literature for the optical warps. It is also not clear whether the onset of warp in the old stellar population follows the Briggs rule. So we would like to understand the onset of warps in the stellar discs; where exactly the warp begins in the disc? What can we learn from the IR observations? Since the onset of warp is global in nature, how is it connected to the baryon distribution as well as the dark matter distribution in disc galaxies? In the present paper, we focus on the issue of the onset of warps in edge-on spiral galaxies based on the deep 4.5 micron {\em Spitzer/IRAC} observations.   

Several mechanisms have been put forward to explain the origin of the classical integral sign 'symmetric' warps in the galactic discs: intergalactic magnetic fields (Battaner et al. 1991), cosmic infall (Jiang \& Binney 1999; Shen \& Sellwood 2006), angular momentum misalignment between the dark matter halo and the disc (Debattista \& Sellwood 1999), and intergalactic accretion flows (Lopez-Corredoira et al. 2002; Sanchez-Salcedo 2006). However, none of them seems to be unquestionable. Here we concentrate on the model where warps are formed by tidal interactions only. In the hierarchical structure formation scenarios, galaxy interactions are common and tidal interactions with nearby companions or satellites have often been considered as a plausible mechanism for the generation of warps in disc galaxies (Zaritsky \& Rix 1997; Reshetnikov \& Combes 1998; Schwarzkopf \& Dettmar 2001; Bailin \& Steinmetz 2003). The principle of warp generation is quite straightforward and generic in the tidal interaction scenario, because the m=1 Fourier component of the perturbing potential is proportional to $\cos(\varphi)$, which can naturally induce an integral sign symmetric warp in the galactic midplane (Saha \& Jog 2006, hereafter SJ06a). We will check our observations against this model only as it makes testable predictions for the onset radius of a warp, which is not trivially the case for other models of warp generation. 

Since warps are observed in both the young stellar populations (Drimmel et al. 2000) and old stellar populations as well as in the neutral hydrogen gas, it is no longer just a gaseous phenomenon. Warping emerges as a self-gravitating phenomenon of the galactic disc. Based on the fact that warp is self-gravitating, SJ06a have recently put forward an explanation for the onset of warps in the galactic discs by calculating the self-consistent linear response of the disc due to tidal interaction. It is possible that warps and their onset radii can be explained by many other mechanisms as mentioned above, but at present we have only SJ06a the model that provides the onset radii for the warps. Building new models for the onset radii under other mechanism is beyond the scope of the present paper.   

Van der Kruit (2007) has made a comparative study of the edge-on galaxies in the Garcia-Ruiz et al. (2002) sample and tried to find the connection between the onset of HI warps and the truncation in the stellar discs. Surprisingly, little attention is given so far to the onset of optical/IR warps predominantly due to the difference in resolution between radio and optical data. It is not clear whether the onset of HI warp coincides with that of the optical warps. The onset of warps might be able to provide valuable information about the structure of disc galaxies. Given the mass model, which can be derived from the photometry and the observed rotation curve of a galaxy, SJ06a allows us to calculate the linear response function of the disc to tidal interaction and derive semi-analytically a minimum radius for the onset of warps. The beauty of the linear response theory lies in that it does not need the distance to the tidal perturber in evaluating the onset of warp. 

Our primary aim in this paper is to measure the onset radii for the warps in a sample of edge-on spiral galaxies observed in the {\em Spitzer/IRAC} 4.5 micron band and then test whether these observed onset radii are consistent with the theoretically estimated onset radii of warps according to SJ06a. Our analysis shows that the onset of warps are generally not symmetric about the center of the galaxy and this fact has led us to investigate the underlying symmetry of the light distribution in edge-on disc galaxies. We then study the correlation between the asymmetry in the onset of warp and the asymmetry (if any) of the light distribution in the sample galaxies. We also study the possible connection between the onset radii for the warps and the dark matter content within the extent of the stellar disc. 

In \S 2 and \S 3, we discuss the sample selection, the modeling of the light distribution and its asymmetries. In \S 4 and \S 5, we develop the basic ingredients for the rotation curve modeling and the disc response to the tidal interaction respectively. \S 6 is devoted to the onset of warps. The comparison between the theory and observation and important results from our analysis are layed out in \S 7. Discussion and conclusions are presented in \S 8. Appendix A presents detailed results on each of the sample galaxies.



\section{\bf{Sample selection and observations}}

The {\em Spitzer/IRAC} observations of edge-on spiral galaxies are from both a dedicated survey (Proposal 20268, ''The Formation of Dust Lanes in Nearby, Edge-on Disk Galaxies'', P.I: R. S. de Jong), and all public {\em Spitzer/IRAC} observations of very high inclination spiral galaxies in the archive that are near enough to easily resolve their vertical disc structure. The sample spans a range of spiral galaxy subtypes and a range of masses. 

{\em Spitzer} data is processed by a pipeline after transmission from the satellite. This pipeline corrects for known behavior of the Infrared Array Camera (IRAC) and takes care of flatfielding the images. Our data was processed with the S13 version of the pipeline or better. The resulting Basic Calibrated Data (BCD) can be retrieved with {\it Leopard}, the online data-retrieval tool. These BCDs were subsequently combined into mosaics with the MOPEX package, the mosaicing software specific for {\em Spitzer/IRAC} data, using input settings kindly provided by Karl. D. Gordon. The resulting mosaics are of similar quality than those produced by the {\em Spitzer Infrared Nearby Galaxy Survey} (SINGS, Kennicutt et al 2003), which used a custom mosaicing pipeline. Remaining issues in the mosaics are leaks of flux from bright stellar objects (mux-bleed), column-to-column background level changes (optical/electronic banding), lower column values due to a bright object (column pull-down) and ghost afterimages of bright objects (persistence). However, the careful dithering of the observations for our dedicated survey has substantially reduced the banding and persistence issues in our data. In possible future versions of the IRAC BCD pipeline or MOPEX, these remaining issues will be taken care of. However, most are small scale and only near bright stars and therefore do not affect the results presented here. The details of our data reduction are described in detail in Holwerda et al. {\em in prep.}

Initially we started with 24 edge-on galaxies. Of these, we found 14 galaxies showing warps (by eye) in channel 2, i.e., the 4.5 micron band of {\em Spitzer}. But 4 galaxy images have some problem, e.g., ESO373 and NGC 5946 both suffer from incomplete coverage on one side and NGC 4437 and NGC 4631 have too mild warps to measure. Our final sample consists of 10 galaxies for which we have a reliable measurement of warps. Only 6 of those galaxies were found to have rotation curves in the literature, which are necessary for our analysis. 

\section{\bf{Modeling the light distribution}}

Before performing our 2-dimensional fit of the edge-on disc in channel 2 (4.5 $\mu$m), we rotated the image mosaics so that the major axis are aligned with the x-axis and degraded the resolution to that of the 8 micron mosaics. We furthermore masked all fore and background objects and other structures --artifacts and bright disc sources-- with a combination of a source extractor (Holwerda 2005) catalog and a visual inspection.

We fitted a projected exponential disc, adapted from van der Kruit and Searle (1981), but with an exponential decline instead of a $sech^2$ in the vertical (z) direction: $\mu(R, z) = \mu(0,0) ~ \left({R \over R_{d}}\right)K_1\left({R \over R_{d}}\right) ~ e^{-{|z|}/z_0}$. The fit is performed with {\it mpfit2dfun} routine using IDL with six parameters: the central position ($x_0$,$y_0$), position angle ($pa$) and disc's central surface brightness ($\mu(0,0)$), scale lengths ($R_d$) and height ($z_0$). The data-reduction and 2D fits to all channels is presented in Holwerda et al. (in prep).

\subsection{Asymmetry in the light distribution of edge-on galaxies}
Apart from the modeling of the radial and vertical variations of the stellar light in the edge-on disc galaxies, we also investigate the underlying symmetry of the light distribution about the centre of the galaxy. There exist many dynamical signatures of asymmetry in the edge-on disc galaxies, e.g. H{\footnotesize I} line-width profiles (Richter \& Sancisi 1994); rotation curves (Sofue 1996; Jog 2002); asymmetric warps (Garcia-Ruiz et al. 2002; Saha \& Jog 2006b), etc. We also find asymmetry in the location of the onset of warps on either side of the disc about the centre as presented in \S 6.1. Are the underlying asymmetries in the light and hence stellar mass distribution causing the asymmetries in the onset of warps? Or are the dynamical asymmetries mainly caused by asymmetric dark matter distribution? These questions have provided the basic motivation to come up with a method to measure the asymmetry in the light distribution of an edge-on galaxy.

To quantify the asymmetry in the light distribution, we use horizontally rotated 4.5 micron images of edge-on galaxies from our sample (which are also being used to measure the radii for the onset of warps, see \S 6) and sum all the flux in two equal-sized boxes encompassing one half of the galaxy, one on the left of the galaxy centre and one on the right. These fluxes on the left and right of the galaxy centre are denoted by $F_{L}$ and $F_{R}$ respectively. 

We then define the asymmetry in the light distribution of edge-on galaxies as a dimensionless and normalized quantity called the 'edge-on asymmetry' ($\alpha_{F}$) as 

\begin{equation}  
\alpha_{F} = \frac{(F_{L} - F_{R})}{(F_{L} + F_{R})}
\end{equation}
  
It is obvious from the above definition that $\alpha_{F} = 0$ refers to a symmetric light distribution about the center of the edge-on galaxy. Note that $\alpha_{F} < 0$ would imply a deficiency in the light on the left side of the galaxy centre compared to that on the right side and vice-versa for $\alpha_{F} > 0$. $\alpha_{F}$ calculated in this way gives us a global measure of the asymmetry in edge-on disc galaxy. Since we use the total integrated flux to calculate the $\alpha_{F}$ value, we loose any local information of asymmetry in the edge-on image of the galaxy. However, it may be interesting to see the variation of $\alpha_{F}$ along the major axis of the galaxy by considering small sized rectangles on either side of the centre of the galaxy. We discuss the possible connection of the edge-on asymmetry with the onset of warps in \S 7.3. 
      
\section{\bf{Galaxy mass models and the rotation curves}}
We consider a three component galaxy model in modeling the observed rotation curves. The galaxy is composed of a bulge, a disc, surrounded by a spheroidal dark matter halo. Below we describe each component in detail.

\subsection{\bf{Bulge model}}
The bulge dominates, in general, the inner parts of the rotation curve. One galaxy in our subsample, NGC7814, is actually bulge-dominated and indeed the bulge is a major contributor to the overall rotation curve of NGC7814. For simplicity, we assume the density profile of the bulge to be described by a two-parameter Plumer-Kuzmin model (Binney \& Tremaine, 1987):

\begin{equation}
\rho_{b}(r)=\frac{3M_{b}}{4\pi R_{b}^{3}}\left(1+\frac{r^{2}}{R_{b}^{2}}\right)^{-5/2},
\end{equation}

\noindent where R$_{b}$ is the bulge scale-length and M$_{b}$ is the total bulge mass.
The potential, $\Phi_{b}$, due to the bulge is obtained by inverting the Poisson equation for the density profile (eq.[2]) and is given in Binney \& Tremaine, 1987 (eq.[2.47a]). We use this potential to derive the bulge contribution to the observed rotation curves.

\subsection{\bf {Stellar disc model}}
The almost negligible influence of dust in the {\em Spitzer/IRAC} (4.5 micron) band allows us to do an accurate 3D mass modeling of the stellar disc. The light distribution of the stellar disc is reasonably well modelled by a double exponential profile. We assume the stellar disc to be axisymmetric and the density distribution is given by:
 
\begin{equation}
\rho_{d}(R,z) \: = \: \rho_{d0}e^{-R/R_d}e^{-{|z|}/z_{\circ}}, 
\end{equation}

\noindent where $\rho_{d0}$ is the disc central density and $R_d$ is the disc scale-length. $z_{\circ}$ is the vertical scale-height that in general is a function of the galactocentric distance $R$. Interactions and mergers are some of the possible candidates that could give rise to flaring discs, i.e., the scale-height $z_{\circ}$ increasing systematically with the radial distance. Apart from the external perturbations, it is possible that some internal disc instabilities could cause such a scenario giving rise to flaring. A number of mechanisms such as scattering due to spiral arms and GMCs (Spitzer \& Schwarzschild 1953; Lacey 1984), levitation (Sridhar \& Tauma 1996), accretion of small satellites (Sellwood et al. 1998), Clustered star formation (Kroupa 2002), unstable bending waves (Sotnikova \& Rodionov 2003) could heat up the disc in the vertical direction but it is not obvious what internal mechanisms could produce radially dependent heating resulting in flaring in the disc. The choice of the double exponential disc comes from a recent study by Schwartzkopf \& Dettmar (2001) of tidally-triggered vertical disc perturbations performed on a fairly large sample of disc galaxies including both interacting as well as non-interacting galaxies. Their study reports that $33\%$ of the discs show an exponential mass distribution in the vertical direction indicating that many disc galaxies have such non-isothermal vertical profiles. Another study by de Grijs et al. (1997a), based on a fairly large number of highly inclined disc galaxies, shows preference towards exponential vertical profiles as compared to the $sech$ or $sech^2$ vertical profiles also proposed for the vertical light distributions (van der Kruit 1988). By solving the hydrodynamical equations incorporating simple star formation law, Burkert \& Yoshii (1996) found that the proto-galactic disc naturally settles down to the vertical exponential stellar distribution. So in the present work, we concentrate on investgating how these non-isothermal (double exponential) discs respond to tidal interaction due to an external perturber. Nevertheless, we will discuss qualitatively the expected differences when adopting a $sech^2$ law for the vertical density distribution instead of an exponential law.

The potential, $\Phi_{d}$, due to the above form of the double-exponential disc is given by Kuijken \& Gilmore (1989) and Sackett \& Sparke (1990):

\begin{equation}
 \Phi_{d}(R,z) \:=\: -4\pi G{\rho_{d0}}{R_{d}}^2 z_{\circ} {\int_{0}^{\infty}}{\frac{J_{0}(k R)}{[1+(k R_{d})^2]^{3/2}}} {\frac{ {{k z_{\circ}}e^{-|z|/z_{\circ}}} - {e^{-k |z|}} } {k^2 {z_{\circ}}^2 -1}}dk
\end{equation}


\subsection{\bf {Dark matter halo model}}

The dark matter halo is modelled by an axisymmetric spheroidal system with a pseudo-isothermal density profile (de Zeeuw \& Pfenniger 1988):

\begin{equation}
 \rho_{h}(R,z) \:=\: \frac{\rho_{h0}}{1 + (R^2 + \frac{z^2}{q^2})/R_{c}^2},
\end{equation}
 \noindent 
where $\rho_{h0}$ is the central mass density of the halo, $R_c$ is the core radius and $q$ is the halo flattening (or oblateness) parameter. Note that $q=1$ would give rise to a spherical halo, while $q=c/a < 1$ produces an oblate halo and $q=c/a > 1$ leads to a prolate halo. Here $a$ is the axis in the disc plane and $c$ is along the vertical direction. The mass density goes as $R^{-2}$ at large radii ($R/R_c \gg 1$) and the mass within a spheroid $M(R) \propto R$ giving rise to an asymptotically flat outer rotation curve.

The calculation of potential ($\Phi_{h}$) due to the dark matter halo is cumbersome and we use eq.(2.88) of Binney \& Tremaine (1987) to calculate the circular velocity in the disc midplane (which in our model is coplanar with equatorial plane of the dark matter halo) for the assumed halo profile.

\begin{table}
\begin{center}
\caption[]{Parameters for the dynamical mass models of the sample galaxies.\label{tbl-2}}
\begin{tabular}{crrrrrrrr}\hline 
Galaxy & \multicolumn{2}{c}{Bulge} & \multicolumn{2}{c}{Disk} &  \multicolumn{3}{c}{Dark matter halo} &M$_h$/M$_d$ \\ 

 &  R$_b$ & M$_b$ & R$_d$ & M$_d$ & R$_c$ & $\rho_{h\circ}$ & $q$ & \\
 & (kpc) & ($\times 10^{10}M_{\odot}$) & (kpc) & ($\times 10^{10}M_{\odot}$) & (kpc) & ($M_{\odot}/pc^3$) & & \\
\hline 
ESO121-G6 &0.90 & 0.10 & 1.52 & 1.08  & 7.6 & 0.01 & 0.95  & 4.85 \\

NGC4013 & 0.70 & 1.20 & 1.63 & 2.3 & 5.05 &0.026 &0.80 &3.56 \\

NGC4157 & 1.10 &0.05 & 1.56 & 2.7 &2.53 &0.093 &0.98 & 3.40\\ 

NGC4302 & 2.50 & 1.60 & 2.88 &  2.2 & 4.61 &0.026 &0.80 & 7.06\\ 

NGC4565 & 0.95 & 2.97 & 4.17 & 8.5 & 8.97 & 0.016 & 0.95 & 5.63\\
  
NGC5907 & 0.60 & 1.30 & 3.72 & 6.8 & 4.84 & 0.050 &0.95 & 6.63\\  
\hline 
\end{tabular}

\end{center}
\end{table}

\begin{table}
\caption[ ]{Basic properties for each galaxies and the dynamical $M/L$ in the {\em Spitzer/IRAC} 4.5 micron band}
\begin{flushleft}
\begin{tabular}{|c|c|c|c|c|} \hline  
Galaxy    & distance$^*$   &  $\mu_{0,4.5micron}$  &  L$_{4.5}$    &  (M/L)$_{4.5}$ \\
        & (Mpc) & (mag/arcsec$^{2}$) & ($\times 10^{10} L_{\odot}$) & (M$_{\odot}$/L$_{\odot}$) \\
\hline 
ESO121-G6  & 17.30  & 15.20 & 2.16 & 0.55  \\

NGC4013  & 11.91  &  14.64 & 3.55 &  0.98  \\ 

NGC4157  & 10.97  &  14.91 & 3.41 &  0.80  \\

NGC4302  & 16.80  & 15.13 & 5.87 &  0.65  \\

NGC4565  & 11.75  &  14.76 & 16.40 &  0.69 \\

NGC5907  & 13.49  &  15.00 & 10.33  &  0.78 \\
\hline 
\end{tabular}
\end{flushleft}
$^*$ Obtained from HyperLeda\\
\label{param}
\end{table}

\subsection{Rotation curves}
For each galaxies in our sample, we have the observed rotation curves either from the H$_\alpha$, CO or H{\footnotesize I} 21 cm observations. We model the observed rotation curves using a bulge, a disc and a dark matter halo.

We assume that the disc and the spheroidal components are aligned with the symmetry axis of the dark matter halo and also assume the virial equilibrium in the galaxy. The total circular velocity in the disc midplane (z=0) can then be written as: 

\begin{equation}
V_{c} = \sqrt(V_{bulge}^2 + V_{disc}^2 + V_{dmh}^2),
\end{equation}

where

$V_{bulge}^2 = R\left.{\frac{\partial \Phi_{b}}{\partial R}}\right|_{z=0}$;          $V_{disc}^2 = R\left.{\frac{\partial \Phi_{d}}{\partial R}}\right|_{z=0}$, and  $V_{dmh}^2 = R\left.{\frac{\partial \Phi_{h}}{\partial R}}\right|_{z=0}$.

\bigskip

Generally speaking, modeling of the rotation curve is a non-linear regression problem in a multi-dimensional parameter space. For example, in our case the number of parameters involved is 9 (2 from bulge, 3 from disc and 3 from dark matter halo + $M/L$) and so the modeling may naturally suffer from the uniqueness problem (Dutton et al. 2005). However, many of these parameters are derived from the analysis of the observed light distribution and hence the number of free parameters is strongly reduced. For example, the observed scale length of the baryon distribution in the galaxy largely reduces the freedom of changing other components to match the observed shape of the rotation curve.

We model the observed rotation curves under the maximum stellar mass assumption, i.e. we assign the maximum mass to the bulge and disc components without overpredicting the rotation curves. From this, we derive the stellar mass (equal to the sum of disc and bulge mass) to light ratio, $M/L$, in the {\em Spitzer/IRAC} 4.5 micron band. The derived $M/L$ values are in reasonable agreement with Bell \& de Jong (2001) type stellar population model predictions. Using their IMF normalization, we expect $M/L$ values of $0.6 - 0.75 M_{\odot}/L_{\odot}$. See Table 1 and 2 for details of the mass models. The mass of the dark matter halo ($M_h$), as appearing in Table 1, is calculated within a radius of 10 disc scale lengths and this notion is used consistently through out the paper. 

\section {\bf {Disk response under tidal interaction} }

We study here the self-consistent linear response of each galaxy assuming that they have gone through some tidal interaction with neighbouring galaxies or satellites in the past. It is true that the warps in our sample galaxies could arise due to other reasons too, e.g., intergalactic accretion flows onto the disc (Lopez-Corredoira et al. 2002), angular momentum misalignment between the dark halo and the disc (Debattista \& Sellwood 1999) or internal mechanisms such as bending instability, etc. We preferred to choose tidal interaction here because this is a likely mechanism (given the fact that interaction with companion is common) and can produce a large-scale warp. A tidal interaction can generate a smooth integral sign symmetric large-scale warp of a galactic disc in a natural way. However, we are mostly interested here in understanding the response of the disc in the linear regime which normally depends on the internal structure of the disc alone as long as the forcing term remains similar. It would be interesting to check the nature of disc response under other mechanisms such as angular momentum misalignment or intergalactic accretion flow, but building models that predict the location for onset of warp under such mechanisms is beyond the scope of the present paper. So based on our original assumption the derived linear response of the disc, which we refer to as warped-response of the disc, shows how a disc galaxy would behave under tidal interaction (tidal force being $\propto cos(\varphi)$, where $\varphi$ is the azimuthal coordinate in the cylindrical geometry assumed for the disc.). The minimum of the warped-response curve, which depends on the internal structure of each galaxy, indicates where the warp begins in the disc. The basic methodology for calculating the warped-response is layed out in detail in SJ06a.

According to the linear response theory, the response potential of the disc $\Psi^{warp}_{resp}$ due to any tidal interaction is directly proportional to the perturbing potential $\Psi_{tidal}$ and is given by the following simple relation:

\begin{equation}
{\Psi^{warp}_{resp}}(R,\varphi,z=0) \:=\: {\mathcal R}_{warp}(R){\Psi_{tidal}(R,\varphi,z=0)},
\end{equation}   

where the dimensionless warped-response of the double-exponential disc embedded in the dark matter halo is given by:

\begin{equation}
{\mathcal R}_{warp}(R) \:=\: -{4 G \rho_{d0}}{ {\int_{0}^{\infty}}} { e^{-{R^{\prime}}/R_d}} {\frac{h(R^{\prime})}{z_{\circ}}} {\mathcal Z}^{warp}_{exp}(R,R^{\prime},0,1) R^{\prime} dR^{\prime}.
\end{equation}

In the above equation, the functions ${\mathcal Z}^{warp}_{exp}$ and $h(R)$ are taken from eq.[20b] and eq.[22b] of SJ06a respectively. The function $h(R)$ describes the shape of the integral sign warp. We use the above equation for generating the warped-response as a function of the galactocentric radius and find the minimum of this response function for each galaxy in our sample. Because of the linear response theory, the minimum of this warped-response does not depend on location of the tidal perturber in the sky.

\section{\bf{Measuring the onset of warps}}

Warping is a global phenomenon of the edge-on galaxies. Since the graininess in the disc is at the level of a radial epicyclic scale, galactic discs can be assumed to be a smooth continuous medium when studying its global bending (whose length scale is compared to the size of the galactic disc itself.). As a result, we expect the warping of the disc to be described by a continuous curve rather than a wiggly discontinuous curve. The warp curve here is an adequate representation of the actual warp of the edge-on galaxy projected in the plane of the sky. Because of the continuous nature of the warp curve, it is not easy to set up a unique criterion for its onset in the galactic disc. We try to treat each galaxy on a general platform and derive the radius for the onset of warp by applying a general principle as outlined below. 

\noindent An almost perfect edge-on galaxy is termed as warped when the deviation of a significant portion of its outer parts is systematically higher with respect to an inner undisturbed plane of symmetry. We consider this inner part of the disc to be flat and determine the point at which the disc bends systematicaly away from it (Jimenez-Vicente et al. 1997; Garcia-Ruiz et al. 2002). We will call the distance of this point from the center of the galaxy image as the onset radius of the warp. Note that in this way, we can have two different radii for the onset of warp on either side of the center in the projected image. 

\noindent The first step in the process of determining the radius for the onset of warp is to select a rectangular slice large enough to encompass the whole image of the edge-on galaxy which is already rotated to be horizontal. We then run this slice of image through a spatial median filter to remove small scale structures. Using the {\it{FIND-GALAXY}} routine (written by Michele Cappellari) we determine the center of the filtered slice of the image. Since this routine also gives us the orientation of the image, we further rotate the image (using IDL code {\it{ROT}} which can rotate the image by any arbitrary angle) by the residual angle to make it almost perfectly horizontal. Once this is done, we fit gaussians using the IDL routine {\it{mpfitpeak}} (written by C. B. Markwardt) perpendicular to the major axis of the galaxy and determine the centroids and their uncertainties at each galactocentric radii. The curve passing through the centroids is what we call the warp curve. Now, we divide the outer parts of the curve into $N$ (typically N $\ge$ 4)  segments on either side of the assumed flat portion and calculate the ratio of the standard deviation of each segment with respect to the flat portion and determine {\it the radius} at which the ratio of the standard deviations increases reasonably systematically on the either side of the inner flat portion. We call this radius the onset of warp ($R_{w}$). This way, we get two radii for the onset: $R^{l}_{w}$ on the left side and $R^{r}_{w}$ on the right side of the image. For some galaxies, the above scheme of determining the onset radii for the warps can fail. This can be the case when there is small scale corrugations and use of eye estimation is preferred. Since our theoretical model is based on the assumption of axisymmetry  and provides only one radius for the onset of warp, we calculate the average radius ($R^{obs}_{w}$) for the onset from the observation in order to compare with the theory:

\begin{equation}
R^{obs}_{w} = \frac{(R^{l}_{w} + R^{r}_{w})}{2}.
\end{equation}

\begin{table}
\begin{center}
\caption[ ]{Measurement of the onset radii of warps in the sample of edge-on galaxies}
\begin{flushleft}
\begin{tabular}{cccccccc}  \hline 
Galaxy    & $R_d$   &  $R_{w}^l$  &   $R_{w}^r$    &  $R^{obs}_{w}$ & $\epsilon_{on}$ & $\alpha_{F}$ \\
               & $('')$     &  $('')$  &          $('')$      &     $('')$    \\
\hline
ESO121-G6  &  18.11  &  96.75 &  84.00 &  90.37 & 0.071 & -0.034$\pm$ 0.0012    \\

NGC4013  &  28.20  &  123.00 &  129.00 &  126.00 & -0.024 & -0.020$\pm$ 0.0047     \\ 

NGC4157  &  29.40  &  145.50 &  135.75 &  140.60 & 0.035 & -0.014$\pm$ 0.0076     \\

NGC4217  &  31.9  &  111.75 &  98.25 &  105.00 & 0.064 & -0.032$\pm$ 0.0012    \\

NGC4244  &  89.35  &  242.25 &  264.75 &  253.50 & -0.044 & 0.014$\pm$ 0.020     \\

NGC4302  &  35.40  &  121.50 &  115.50 &  118.00 & 0.025 & -0.0152$\pm$ 0.0031     \\

NGC4565  &  73.25  &  405.75 &  378.75 &  392.25 & 0.034 & -0.027$\pm$ 0.0037    \\

NGC5907  &  56.90  &  233.25 &  228.00 &  230.90 & 0.012 & 0.0024$\pm$ 0.0022    \\

NGC7090  &  32.55  &  91.50 &  82.50 &  87.00 & 0.052 & -0.094$\pm$ 0.0042   \\

NGC7814  &  23.06  &  135.00 &  121.50 &  128.25 & 0.053 & -0.053$\pm$ 0.0059   \\
\hline
\end{tabular}
\end{flushleft}
\label{param}
\end{center}
\end{table}


 \subsection{Onset-asymmetry} 
            
A new quantity of interest emerging from our analysis is the asymmetry in the onset of warp in the disc. We call this the {\it{onset-asymmetry}} of the warp and quantify it in terms of a dimensionless quantity called onset-asymmetry index ($\epsilon_{on}$):

\begin{equation}  
\epsilon_{on} = \frac{(R^{l}_{w} - R^{r}_{w})}{(R^{l}_{w} + R^{r}_{w})}.
\end{equation}

According to this definition, we have $\epsilon_{on}= 0$ when $R^{l}_{w}= R^{r}_{w}$, meaning a symmetric onset of the warp. This could arise when there is absolute axisymmetry in the underlying potential of the galaxy. On the other hand, when either $R^{l}_{w}= 0$ or $R^{r}_{w} = 0$, the absolute value of the onset-asymmetry index $|{\epsilon_{on}}|$ becomes unity. $|\epsilon_{on}| = 1$ implies a one-sided warping of the disc. So this onset-asymmetry index (0 $\le |{\epsilon_{on}}| \le $ 1) can capture a fairly wide range of galactic warps found in observations. Now, depending on the structure of the underlying galaxy potential there could be situation where $R^{l}_{w} > R^{r}_{w}$ leading to $\epsilon_{on} > 0$ or vice-versa leading to $\epsilon_{on} < 0$ for the warp. In Table 3, we show measurements of the onset radii of warps for 10 galaxies. We also calculate the onset-asymmetry index for each of these galaxies.
 
Warps normally begin around $4 - 6$ scale lengths in galactic discs (e.g., Briggs 1990). At these radii disc truncation frequently occurs (Van der Kruit 2007) and this is also the region where the dark matter halo begins to dominate the disc dynamics. Since the onset of warp falls in this {\it{``conspiracy zone''}} of the galaxy, it is difficult to pinpoint the cause for the onset-asymmetry. However, as mentioned earlier in this paper that the onset of warp is a non-local problem because it involves the bending of the whole disc itself. So at this moment it is clear that the onset-asymmetry might be due to some asymmetry either in the baryon or in the dark matter distribution or both about the centre of the galaxy. Needless to mention that an accurate estimation of the onset-asymmetry index ($\epsilon_{on}$) is potentially important to understand the detailed structure of the underlying galaxy potential. 

\section{\bf{Comparison between the theory and observations}}

In Table 4, we present our analysis of the warp modeling for the 6 edge-on spiral galaxies for which we have rotation curves. Our observational analysis shows that the onset radii for the warps lie within $\sim 3 - 6$ disc scale lengths which is a bit larger range than the onset radii for the gaseous warps. However, given the small sample of galaxies and the uncertainties in determining the scale lengths, our analysis rather indicates that the radii for the onset of stellar warps and H{\footnotesize I} warps are comparable. This suggests that disc warping may actually be a self-gravitating phenomenon which does not differentiate between the gas and the stars in the disc at least in the linear regime. Of course this should be checked carefully for a bigger sample of galaxies. The self-gravity of the disc plays an important role in deciding the location for the onset of warps. Apart from the self-gravity, the onset of warp in the disc is influenced by two apparently independent quantities: the disc thickness parameter which denotes the 3D structure of the disc and the amount of dark matter within the optical disc. We do not know a priori any correlation between the disc thickness parameter and the dark matter content within the optical disc. It may therefore be hard to disentangle the effect of disc thickness parameter and the amount of dark matter within the optical disc when modeling the onset radius for the warp. This is the reason why we are not able to make any conclusive statement regarding the role of self-gravity alone on the onset of warps in real observations.

\begin{figure}
\rotatebox{-90}{\includegraphics[height=7.0 cm]{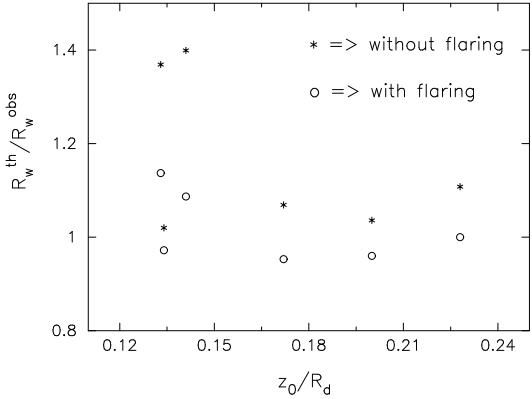}}
\caption{Variation of the ratio of theoretical-to-observed onset radii of warps with disc thickness parameter. The theoretical models with measured flaring (denoted by open circles) are favoured over models without flaring (denoted by stars).  \label{fig1a}}
\end{figure}

\begin{figure}
\rotatebox{-90}{\includegraphics[height=8.0 cm]{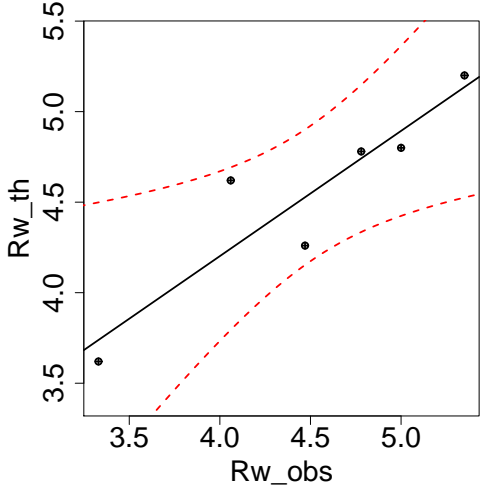}}
\caption{Direct comparison of the onset radii from the observations (along x-axis) and the theoretical models (along y-axis) with measured flaring in the stellar disc. Both axes use unit of a scale length $R_d$. The dashed lines show the $95\%$ confidence band. \label{fig1a}}
\end{figure}

\begin{table}
\begin{center}
\caption[ ]{Comparing the observed and theoretically predicted onset radii}
\begin{flushleft}
\begin{tabular}{ccccccc}  \hline 
Galaxy   & z$_\circ$/R$_d$  &  R$^{obs}_{w}$/R$_d$  &   R$^{th}_{w}$/R$_d$ &  R$^{th}_{w}$/R$_d$ & Flaring rate\\
        &    &            & (without flaring)  &          (with flaring)  & (per scale length) \\
\hline
ESO121-G6 & 0.200  & 5.00 & 5.18 & 4.80  & 0.10   \\

NGC4013   & 0.172  &4.47 & 4.78 & 4.26  & 0.12   \\

NGC4157   & 0.228   &4.78 & 5.34 & 4.78  & 0.14  \\

NGC4302   & 0.141  &3.33 & 4.62 & 3.62  & 0.35 \\

NGC4565   & 0.134  & 5.35 & 5.46 & 5.20  & 0.06 \\



NGC5907   & 0.133 & 4.06 & 5.50 & 4.62  & 0.24 \\


\hline
\end{tabular}
\end{flushleft}
\label{param}
\end{center}
\end{table}

\begin{figure}
\rotatebox{-90}{\includegraphics[height=6.0 cm]{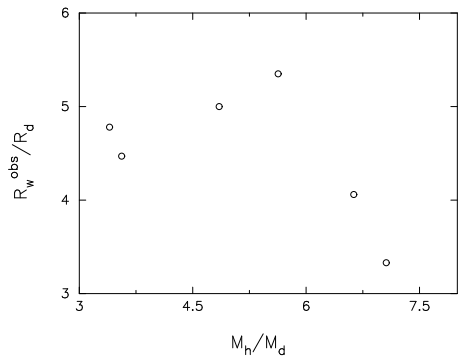}}
\caption{Variation of the onset radii for the warps with the ratio of the dark matter mass to the disc mass calculated within 10 disc scale lengths. The unit of y-axis is in disc scale length ($R_d$). \label{fig1a}}
\end{figure}

\subsection{Onset radii and disc thickness}

In Fig.~1, we show the ratio of theoretically estimated onset radii to the observed ones versus the disc thickness parameter (z$_\circ$/R$_d$). We first reproduce the onset radii for these galaxies using a constant thickness parameter (meaning a disc without stellar flaring) from the analytical calculation. The theoretical predictions in this way overestimates the onset radii for the warps in these galaxies as can be seen from Fig.~1. The reason for this overestimation lies in the assumption of the constant thickness of the galactic disc, causing the disc self-gravity  to increase and consequently the disc to become stiffer. However, by relaxing this assumption and considering the galactic disc to be moderately flaring, the theoretical predictions can match quite well the observed radii for the onset of warps. Such moderate flaring of the galactic discs can possibly arise due to some tidal interactions with nearby galaxies, minor mergers, or satellite infall (Quinn et al. 1993; Schwartzkopf \& Dettmar 2001; Kazantzidis et al. 2008). 

In order to have a proper comparison between the observation and the theory, we measure the rate of flaring in the stellar disc. This is accomplished by fitting gaussian profiles along the vertical direction at each point along the major axis of each galaxy image in the 4.5 micron band and then averaging over both sides of the disc. The reason we use averaged flaring is to have a better comparision with the theory (SJ06a) which is built on axisymmetric models. The measured flaring rate varies from $\sim 0.1 - 0.35$ per scale length (see Table 4) in our sample of galaxies. Note that these flaring measurements are normally an overestimation because we use linear fitting to the flaring curves. Such flaring rate would increase the disc scale height by a factor of $\sim 1.6 - 3.0$ over the optical disc. These values are still, in general, in good agreement with previous observations in other galaxies (see de Grijs \& Peletier 1997; Narayan \& Jog 2002).    

We note that the models with moderate flaring are more favourable to the present observational analysis. In Fig.~2, we carry out a detailed statistical analysis for the model with observed flaring. Fig.~2 shows the direct comparison between the theoretical onset radii calculated incorporating the observed flaring and observed onset radii. As expected, we fit a simple straight line through the data points. The dotted curves are the boundary of a 95\% confidence ellipse and the interior region bounded by the dotted curves is the 95\% confidence band for the true linear regression model. 

\begin{figure}
\rotatebox{-90}{\includegraphics[height=6.0 cm]{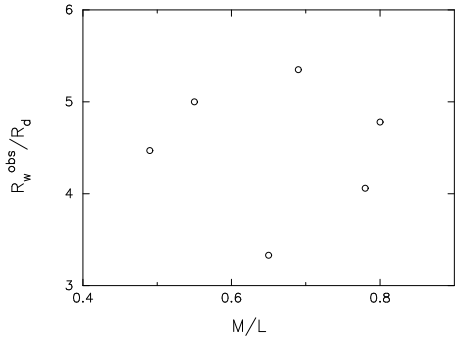}}
\caption{The observed warp onset radii versus the stellar mass-to-light ratio ($M/L$) in the {\em Spitzer/IRAC} 4.5 micron band. The $M/L$ values are in units of solar $M_{\odot}/L_{\odot}$ and the onset radii for the warp are in units of disc scale lengths ($R_d$).}
\end{figure}

\subsection{Dependence of onset radii on dark matter and $M/L$}
We study here the effect of dark matter on the onset radii of the warps. As the dark matter halo becomes more and more massive compared to the disc, it pushes the onset radius inward in the disc as explained in SJ06a (see their Fig.~3). We try to verify the effect of the dark matter in our observational analysis. In Fig.~3, we show the observed onset radii versus the ratio of the dark matter to the disc mass ($M_h/M_d$) derived from the rotation curve modeling. The observed onset radii are in reasonably good qualitative agreement with what had been predicted in SJ06a. Since the analysis of SJ06a is linear, it is hard to find a good match with real observations. The onset radii for the warps decrease as the amount of dark matter increases. Also the theoretical predictions of the onset radii using the model with moderate flaring are in good agreement with the observations. 

In Fig.~4, we aim at understanding the connection between the baryonic $M/L$ and the observed onset radius of the warp of a disc galaxy. According to theoretical expectations, with the increase of $M/L$, the disc mass increases and this in turn increases the self-gravity. As a consequence, the warp should start at a larger radius. However, this trend is not very clear from Fig.~4. For a given rotation curve, the interplay between the $M/L$ and the amount of dark matter is complicated and it is a bit too simplistic to expect such correlation in the first place. It is possible that $M/L$ does not vary much in the Spitzer bands. The observed scatter in the plot may not be surprising given that there can be uncertainties associated with these $M/L$ values. It is of course a non-trivial job to have an accurate error estimation on the $M/L$ values. The primary source of error on $M/L$ comes from the mass modelling which requires velocity profiles, observed luminosities, inclination angles etc. In the present study, the dominant error on $M/L$ is due to the modelling of the rotation curves that are available in the literature. Since none of the rotation curves in our sample comes with error bars, we indeed suffer from accuracy in the modelling and a formal error calculation is not possible here. Instead, we perform an order of magnitude estimate for the errors on $M/L$ values based on the overall fitting of the rotation curves. We used relative root mean square error deviation to get an overall error between the model and the observed rotation curve. The typical model uncertainties derived in this way lie in the range $\sim 10 -  40 \%$. Note that these errors should be regarded as the maximum on the derived $M/L$ values from the fitting procedure. These errors are more or less consistent with the scatter in the $M/L$ values and thus the $M/L$ values are what one would expect from the stellar population models (Bell \& de Jong 2001). This in turn indicates that these galaxies are reasonably well modeled with maximum disc hypothesis.

\begin{figure}
\rotatebox{-90}{\includegraphics[height=6.0 cm]{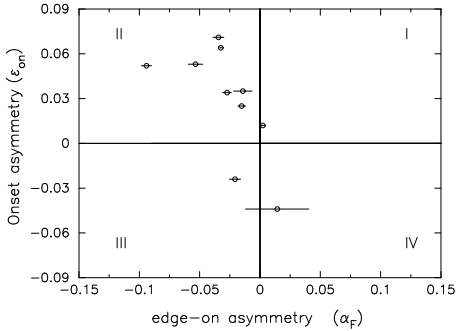}}
\caption{Correlation between the onset asymmetry in warps and the edge-on asymmetry in the stellar light. The horizontal error bars on each measurement of the edge-on asymmetry arise mainly due to the uncertainity in the sky. The galaxies falling in the 1st and 3rd quadrants showing the positive correlation. \label{fig1a}}
\end{figure}

\subsection{Onset asymmetry and edge-on asymmetry}
We measure the onset-asymmetry and the edge-on asymmetry for ten galaxies in our sample as shown in Table 3. As mentioned in \S 3.1 and \S 6.1., the basic aim of these measurements was to see if there is any correlation between the asymmetries which are seen in the onset of warps and the asymmetries in stellar light distribution about the centre of galactic disc. Now, based on a naive theory, we expect the warp should start at a larger radii if the disc self-gravity is dominating over other factors which might be influencing the onset of warp. So in an edge-on galaxy, if there is more light on one side of the centre compared to the other, the warp should begin at a larger radii on that side. This simple idea tells us that if the edge-on asymmetry (for the light) index ($\alpha_{F}$) is positive, it implies a positive onset-asymmetry index ($\epsilon_{on}$) for the warp too (according to our definition). In Fig.~5, we show the variation of onset asymmetry in warps with the edge-on asymmetries in the stellar light. Out of ten galaxies in our sample, only two galaxies (namely NGC 4013 and NGC 5907) show positive correlation and the rest of them show anti-correlation. 
Can we understand this anti-correlation between the two physical entities? The onset of warp is a non-local phenomenon and the disc's self-gravity plays a dominant role in deciding where the warp will begin, irrespective of whether the driving force coming from a tidal perturber (e.g. statellite or companion galaxy), misaligned dark matter halo or any other relevant source. However, any asymmetry in the location of the onset of warp in the disc could likely be a reflection of the asymmetry in the underlying total potential of the galaxy which includes a dominant contribution from the surrounding dark matter halo. Although the light distribution is not symmetric in our sample of galaxies, the total potential arising out of the baryon distribution could still be smooth or the degree of asymmetry in the disc potential is not adequate enough to influence or break the symmetry in the onset of warp. This could be the reason why we do not see any positive correlation in our sample, but still it remains to explain the anti-correlation. Tidal interactions with companion galaxies are common and it is possible that due to such interaction a large scale wake (say an m=1 mode) is set up in the dark matter halo beside a warp in the disc. Such a large scale wake in the dark matter halo could be long lasting (Weinberg 1994) and likely to influence the onset of warp in the disc. Even local asymmetries (e.g. substructures) in the dark matter halo could as well affect the onset radius of warp in the disc. Our speculation that the surrounding dark matter potential is asymmetric and causing the observed anti-correlation is supported by the fact that the inner regions of dark matter halos are seen to be asymmetric in the Millennium simulation (Gao \& White 2006).            

We have also studied the possible correlation of the onset-asymmetry index with other relevant galaxy properties. Fig.~6 shows that the onset-asymmetry index tends to be larger in galaxies with smaller scale lengths of the baryonic mass distribution. Since many galaxies in our sample have comparatively smaller scale lengths, we plan to carry out this analysis in future for a larger homogenoeus sample containing galaxies with equal distribution of scale lengths over a chosen range. However, we do not find any good correlation between the asymmetry index and the core radius of the dark matter halo or the total mass of the halo. Infact, the number of galaxies (6) available for this analyses is too small too draw any conclusive statement. On the other hand, its not even understood how the onset-asymmetry index would depend on these parameters given that there are many other influencing factors in actual observations.  
          
\begin{figure}
\rotatebox{-90}{\includegraphics[height=6.0 cm]{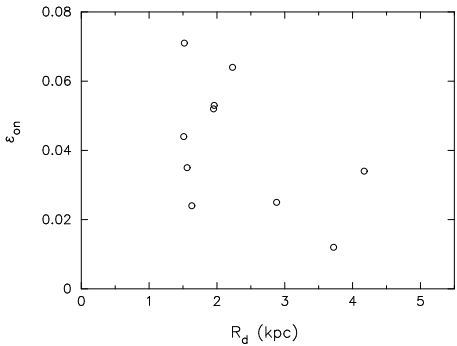}}
\caption{The onset asymmetry index versus the stellar disc scale length ($R_d$). The scale lengths are measured in the 4.5 micron band of the galaxy images.  \label{fig1a}}
\end{figure}


\section{Discussion and Conclusions}

\noindent In this paper, we have studied the nature of the onset of warps in the discs of edge-on spiral galaxies observed by IRAC camera on board Spitzer. Our observational analysis shows that the locations of the onset of warps in the stellar discs have a slightly wider range (3 - 6 $R_d$) as compared to the H$\footnotesize I$ warps as shown by Briggs (1990). By studying the onset of warps, it is possible to shed light on the structure of disc galaxies. Measuring the location for the onset of warp is of geometric nature and through such geometrical measurements, it is possible to infer or constrain some associated dynamical quantities (e.g., dark matter mass within the stellar disc, disc thickness parameter, flaring in stellar disc, $M/L$ etc.) in edge-on galaxies. However, in order to achieve this, we first need to benchmark the relations between the onset radius of warp with those dynamical parameters for a larger number of galaxies. The main important factors which govern the onset of warps in galactic discs are the rate of flaring in the stellar disc and the amount of dark matter within it. However, we note that changing the rate of flaring is more effective in determining the radius for the onset of warps than changing the dark matter mass-to-stellar mass ratio ($M_h/M_d$). The observed onset radii for the warps favour a model with moderate flaring of the galactic disc. 

\bigskip

\noindent In order to model the onset of a warp, we have assumed a double exponential disc to model the light distribution in the vertical direction as well as radial. Near-IR observations of edge-on disc galaxies show that an exponential distribution is a better fit of the observed vertical light distribution than the $sech^2$ distribution (de Grijs et al. 1997b). Still, it is interesting to know how our results would be affected by a different choice of vertical light distribution. Near-IR observations show an excess light over the $sech^2$ model near the galactic mid-plane and thereby a preference towards exponential distribution. In this case, a $sech^2$ distribution would turn the disc to a less massive one over the exponential one (assuming a constant $M/L$ for the galaxy) and thereby its self-gravity would be reduced. We would expect thus to have a smaller radius for the onset of warp for a disc modeled by a vertical $sech^2$ distribution. But it is not possible to give a quantitative estimate without doing a detailed analysis. We intend to model this in more detail in the future.  

\bigskip
\noindent Interestingly, our measurements of the radii for the onset of warps show that warps do not begin from the same radius on either side of the edge-on disc galaxy. Most of the galaxies in our sample show this kind of asymmetry. What causes such asymmetry? Since the warping is a global phenomenon (typical length scale of warp $\sim$ size of the disc itself) and the onset of such a large scale bending is certainly a dynamical process, any asymmetry associated with the onset of warp must be a reflection of the inherent asymmetry in the total galaxy potential. The total potential arises out of the baryon (here, stellar) and dark matter distribution in the galaxy. Therefore, it is possible that the asymmetry in the onset of warp could be either due to the asymmetry in the potential arising out of the baryon distribution or dark matter distribution or both. As a first step towards such a diagnostics, we have measured the global asymmetry in the light distribution of the edge-on galaxies as described in \S 3.1. 

Most galaxies in our sample show an anti-correlation between luminosity distribution on either side of the centre and warp asymmetry (Fig.~5), which indicates that the potential arising out of the baryon distribution asymmetry is not affecting the symmetry in the onset of warps.  
This suggests that the onset asymmetry in warps might be due to the asymmetry in the dark matter potential. In a  hierarchical $\Lambda$CDM universe dark matter halos contain a lot of long-lasting substructure at all scales that could give rise to gravitational potential asymmetries uncorrelated with the baryon asymmetry. Another possiblity is the galaxy-galaxy interaction which can produce long-lasting large-scale non-axisymmetric modes in the dark matter halos. These modes can also be the cause for the asymmetry in the underlying dark matter potential. However, to conclusively say such thing, we need to repeat this excercise for a larger, statistically significant, sample of galaxies.

\bigskip
Our main conclusions are as follows:

1. We provide the measurements for the onset radii of warps for ten galaxies observed in the {\em Spitzer's} 4.5 micron band. We compare six onset radii with the theoretically estimated radii for the onset of warps based on self-consistent linear response theory. The onset radii for the warps lie within 3 - 6 disc scale lengths in our sample of galaxies. We find that the theoretical predictions from the models with a constant disc thickness overestimates the observed onset radii for the warps. Theoretical models with a moderately flaring discs are favoured for explaining the observations. 

2. The observed onset radii for the warps are seen to be declining with the increasing amount of dark matter within the stellar disc (as was shown theoretically in SJ06a). So in a dark matter dominated disc, the warp is expected to be starting at a smaller radius. The location for the onset of warp in a disc does not show any clear correlation with the $M/L$ in our sample of edge-on galaxies.

3. The onset of warps in the discs show a clear asymmetry and we quantify these asymmetries by an index called the onset-asymmetry index. The onset of warps is more likely to be asymmetric in galaxies with smaller scale lengths. 

4. We show that in most of the cases the asymmetries in the onset of warps are anti-correlated with the measured global asymmetries in the stellar light distribution in our sample of edge-on galaxies. We speculate that the onset asymmetries in warps could plausibly arise due to the underlying asymmetry in the dark matter halo potential. The onset-asymmetry index can possibly be used as a potential tool for studying the detailed structure in the dark mater halos.

\bigskip
\noindent {\bf Acknowledgments:}

KS would like to thank the DDRF grant of Space Telescope Science Institute operated for NASA by AURA for carrying out most of the research in this paper. This work is based on observations made with the Spitzer Space Telescope, which is operated by the Jet Propulsion Laboratory, California Institute of Technology under a contract with NASA. Support for this work was provided by NASA through an award issued by JPL/Caltech.  We thank the anonymous referee for valuable suggestions and comments which have improved the quality of the manuscript. KS would like to thank Samir Saha for useful discussions on statistics. We acknowledge the usage of the statistical computing and graphics package R (http://www.r-project.org/). We acknowledge the usage of the HyperLeda database (http://leda.univ-lyon1.fr).


\appendix

\section{Detailed Notes on individual galaxies}
We present here the observational analysis and results for each galaxy in our sample. For each galaxy we show an image, a warp curve, observed and model rotation curves, and the warped response curve using self-consistent linear response theory.

{\bf{ESO 121-G006:}}

This galaxy shows a classical integral sign warp. The disc of the galaxy is slightly thicker than that of our Galaxy as can be seen from the thickness parameter ($a=z_0/R_d$) in Table 4. We use the disc parameters derived from the photometric analysis to model the observed H$\alpha$ rotation curve (Mathewson et al. 1992; Christlein \& Zaritsky 2008) as outlined above in \S 4. The bulge, disc and dark matter halo parameters, derived from the modeling of the observed rotation curve, are shown in Table 1. Note that the dynamical mass of the disc of this galaxy is very small compared to other galaxies in the sample. The disc and dark matter halo parameters are then used to calculate the theoretical radius ($R_{w}^{th}$) for the onset of warps using eq.(8) (see Table 4). A constant thickness disc does not seem to explain the observed onset radius. However, the observed flaring of the disc results in a slight underestimate of the onset radius.
\clearpage

\begin{figure*}
  \newlength{\figwidth}
  \setlength{\figwidth}{\textwidth}
  \addtolength{\figwidth}{-\columnsep}
  \setlength{\figwidth}{0.5\figwidth}
  
  \begin{minipage}[t]{\figwidth}
    \mbox{}
    \vskip -7pt
    \centerline{\includegraphics[width=1.0\linewidth,angle=0]{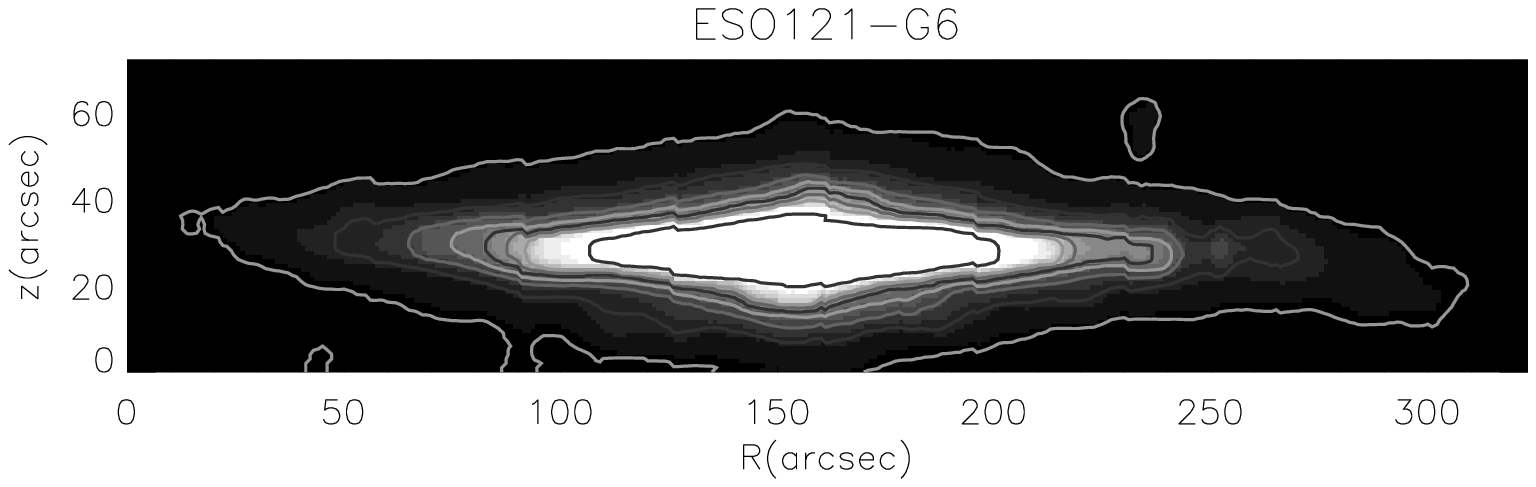}}
  \end{minipage}
  \hfill
  \begin{minipage}[t]{\figwidth}
    \mbox{}
    \vskip -7pt
    \centerline{\includegraphics[width=0.72\linewidth,angle=270]{pgvc_e121.ps}}
  \end{minipage}
\begin{minipage}[t]{\figwidth}
    \mbox{}
    \vskip -7pt
    \centerline{\includegraphics[width=1.0\linewidth,angle=0]{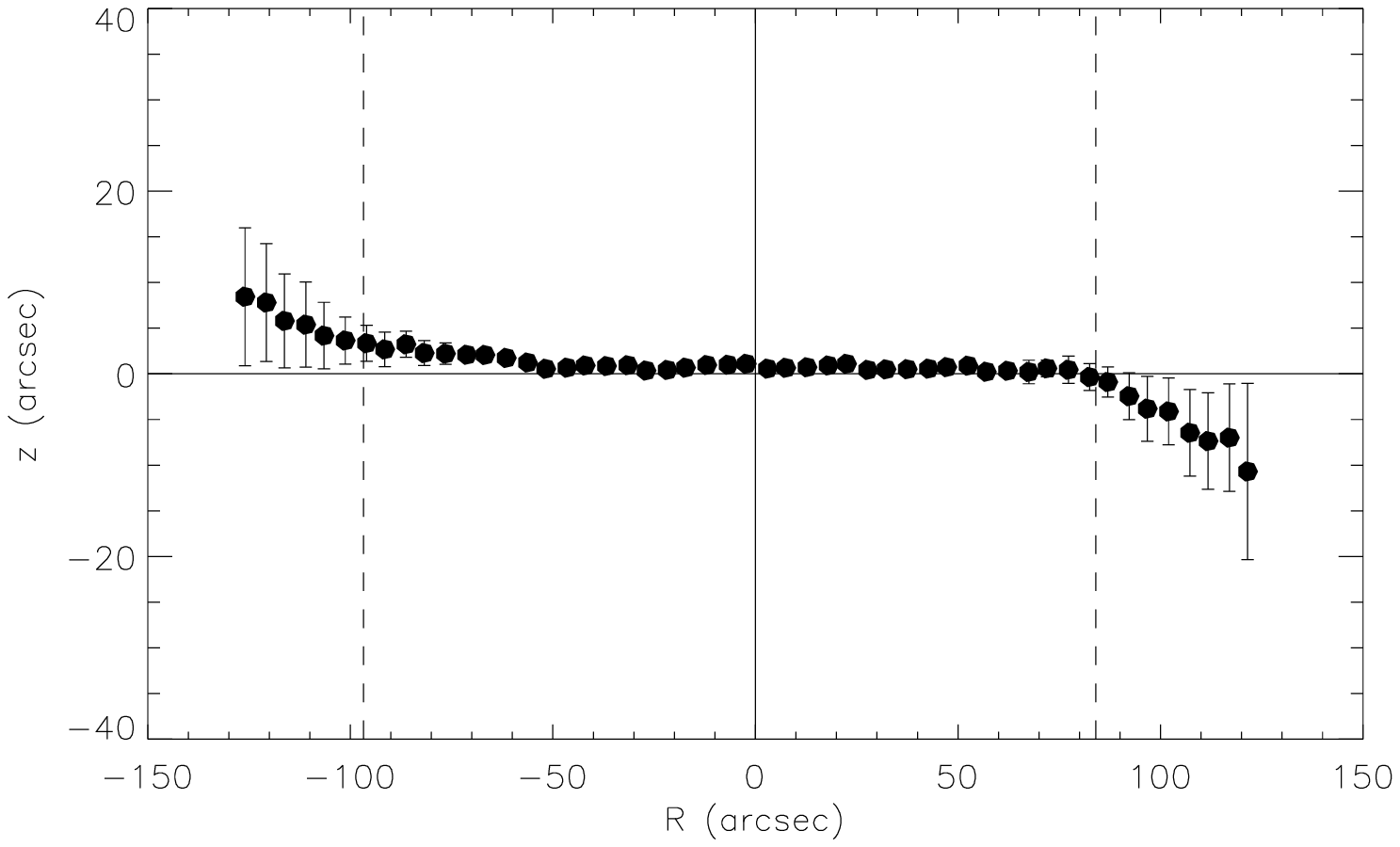}}
  \end{minipage}
  \hfill
  \begin{minipage}[t]{\figwidth}
    \mbox{}
    \vskip -7pt
    \centerline{\includegraphics[width=0.72\linewidth,angle=270]{pgresp_e121.ps}}
  \end{minipage}
    \caption{ (Top left): The iso-intensity contours overlayed on {\em Spitzer/IRAC} 4.5 micron grey-scale image of ESO121-G006. The innermost contour is at $10\%$ of the maximum intensity of the galaxy, subsequent contours are in steps of a factor of two fainter. (Bottom Left): warp curve for the galaxy. The filled circles indicate the vertical offset from the midplane for a given radius, where the error bars denote the 1$\sigma$ uncertainty in the gaussian fitting. The vertical dashed lines indicate the warp onset radii. (Top right): rotation curve for the galaxy. Open circles denote the observed rotation speeds. The dot, dashed and dash-dot lines represent the bulge, disc and dark matter contribution to the rotation curve respectively. The solid line is the net modelled rotation curve for the galaxy. (Bottom right): the plot shows the warped response curve and the minimum of this curve is the onset radius for the warp (see Table 4). The vertical arrow denotes one disc scale length.
 }

\end{figure*}
\clearpage

{\bf{NGC 4013:}}

This is a comparatively smaller sized, perfectly edge-on spiral galaxy lying in the constellation of Ursa Major. NGC 4013 is famous for its prodigious warp, studied in great detail in H$\footnotesize I$ by Bottema (1996). The galaxy shows a mild warping of the disc in the 4.5 micron image as well as in the optical R band observation (Schwartzkopf \& Dettmar 2001). Our rotation curve modeling yields a dynamical mass of the disc that is quite similar to that of Bottema (1996). The shape of the available H$\footnotesize I$ rotation curve (from Bottema 1996) seems to require a rather larger core radius for the dark matter halo. Using the 4.5 micron scale length ($\sim$ 1.63 kpc) for the disc, we get the core radius R$_c$ $\sim$ 5 kpc which is quite different from the previous studies. In the previous study done by Bottema (1996), a larger disc scale length ($\sim$ 2.3 kpc) was used for the rotation curve modeling. The parameters of the galaxy mass model from our analysis are shown in Table 1. By incorporating the observed flaring, we get a smaller onset radius for the warps than the observed one. However, the difference is very small ($\sim 0.2 R_d$). 
    
\clearpage

\begin{figure*}
  \setlength{\figwidth}{\textwidth}
  \addtolength{\figwidth}{-\columnsep}
  \setlength{\figwidth}{0.5\figwidth}
  
  \begin{minipage}[t]{\figwidth}
    \mbox{}
    \vskip -7pt
    \centerline{\includegraphics[width=1.0\linewidth,angle=0]{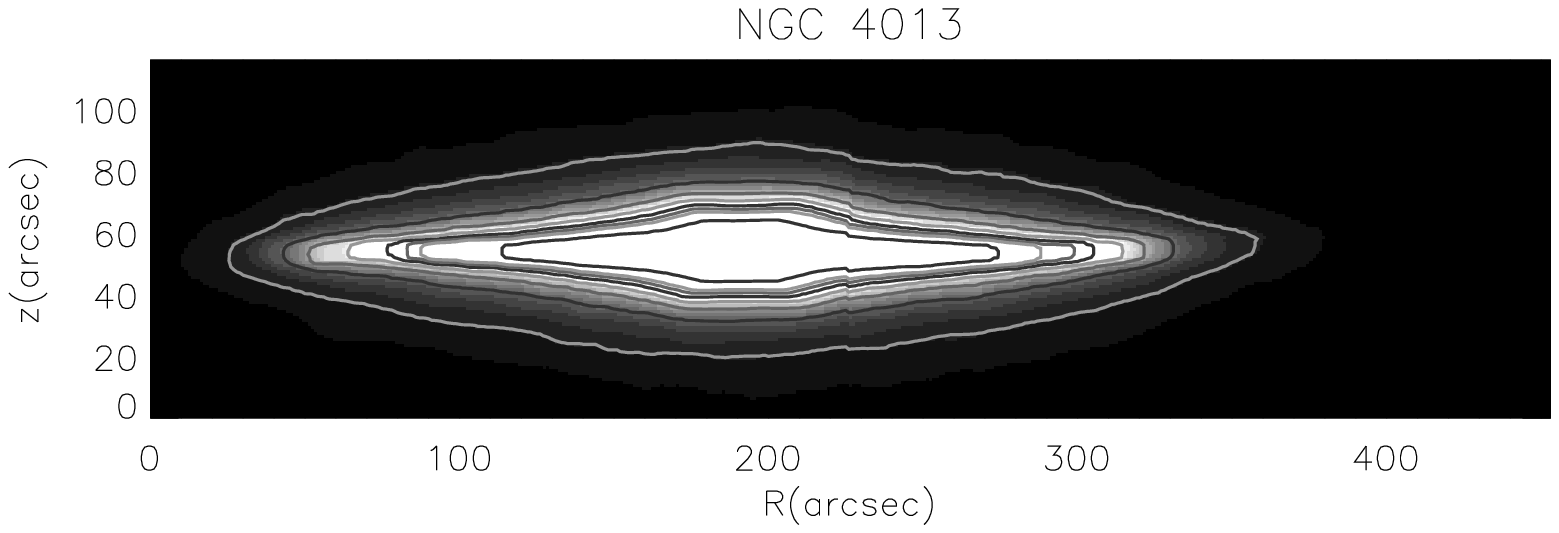}}
  \end{minipage}
  \hfill
  \begin{minipage}[t]{\figwidth}
    \mbox{}
    \vskip -7pt
    \centerline{\includegraphics[width=0.72\linewidth,angle=270]{pg_vc4013.ps}}
  \end{minipage}
\begin{minipage}[t]{\figwidth}
    \mbox{}
    \vskip -7pt
    \centerline{\includegraphics[width=1.0\linewidth,angle=0]{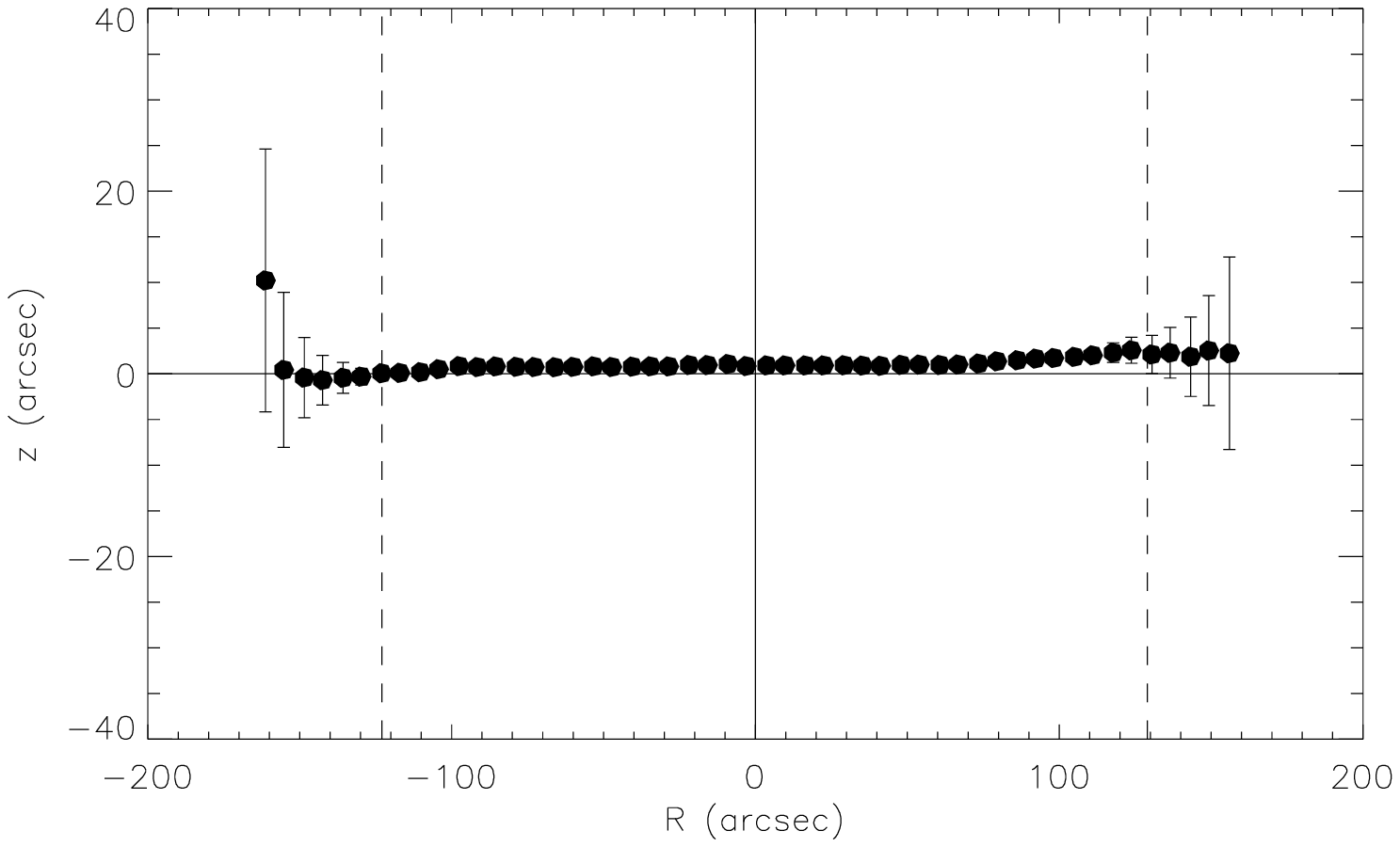}}
  \end{minipage}
  \hfill
  \begin{minipage}[t]{\figwidth}
    \mbox{}
    \vskip -7pt
    \centerline{\includegraphics[width=0.72\linewidth,angle=270]{pgresp_n4013.ps}}
  \end{minipage}
    \caption{Same as Fig.~A1 but for NGC 4013 }

\end{figure*}


\clearpage

{\bf{NGC 4157:}}

NGC4157 is a small barred spiral edge-on galaxy lying in the constellation of Ursa Major roughly in between two big galaxies M106 and M109. Its neighbour NGC4088 is also an edge-on spiral. We detect a small warp in the 4.5 micron band image of the galaxy.  The observed H$\footnotesize I$ rotation curve is obtained from Marc Verheijen (Verheijen \& Sancisi 2001). Note that even though the galaxy is small, it is not dominated by the dark matter halo in our model. Infact, the ratio of dark to disc mass ($M_h/M_d$) is the smallest in the sample. The theoretically estimated onset radius ($R_{w}^{th}$), using the constant scale height for the stellar disc, overestimates the observed onset radius by 0.5 disc scale lengths (i.e. by $\sim$ 10 \%). Using the measured flaring of the stellar disc, we get a perfect match between the theory and observation. 
    
\clearpage

\begin{figure*}
  \setlength{\figwidth}{\textwidth}
  \addtolength{\figwidth}{-\columnsep}
  \setlength{\figwidth}{0.5\figwidth}
  
  \begin{minipage}[t]{\figwidth}
    \mbox{}
    \vskip -7pt
    \centerline{\includegraphics[width=1.0\linewidth,angle=0]{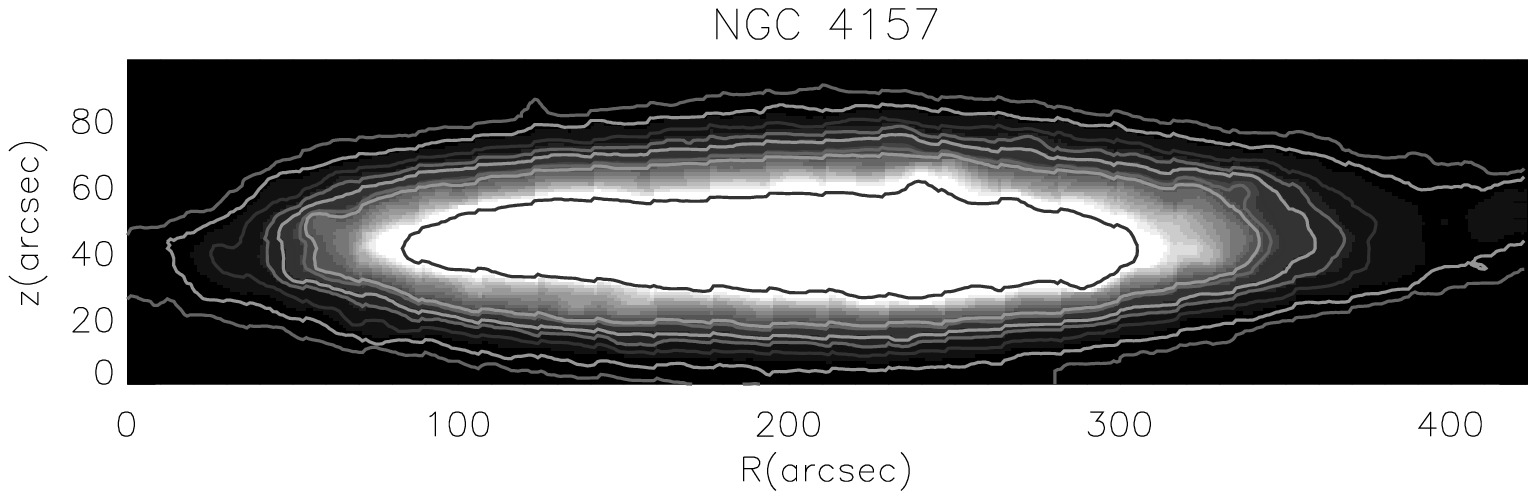}}
  \end{minipage}
  \hfill
  \begin{minipage}[t]{\figwidth}
    \mbox{}
    \vskip -7pt
    \centerline{\includegraphics[width=0.72\linewidth,angle=270]{pgvc_n4157.ps}}
  \end{minipage}
\begin{minipage}[t]{\figwidth}
    \mbox{}
    \vskip -7pt
    \centerline{\includegraphics[width=1.0\linewidth,angle=0]{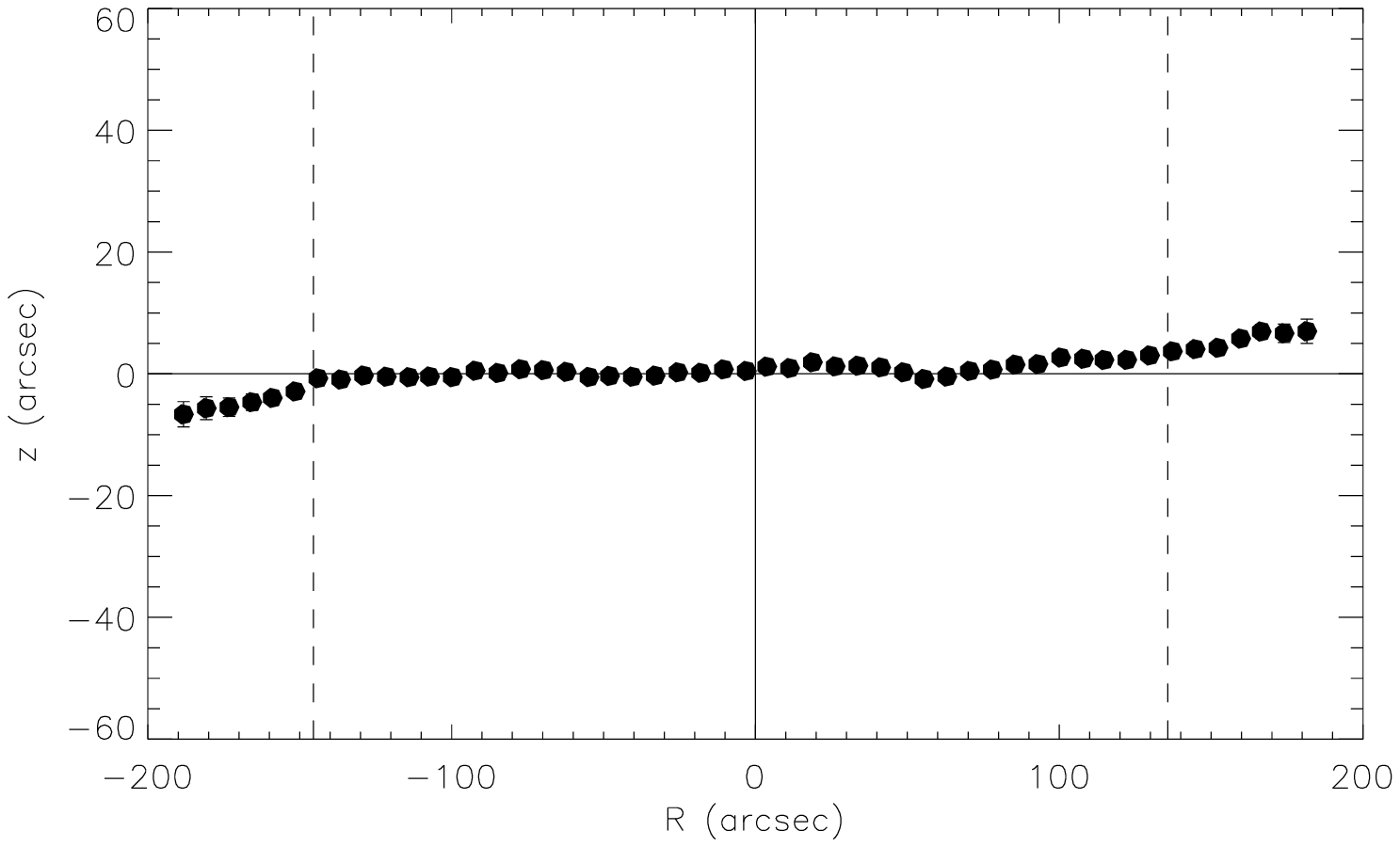}}
  \end{minipage}
  \hfill
  \begin{minipage}[t]{\figwidth}
    \mbox{}
    \vskip -7pt
    \centerline{\includegraphics[width=0.72\linewidth,angle=270]{pgresp_n4157.ps}}
  \end{minipage}
    \caption{Same as Fig.~A1 but for NGC 4157 }

\end{figure*}

\clearpage

{\bf{NGC 4302:}}

NGC4302 is a very nearly edge-on spiral galaxy lying in the constellation of Coma Berenices and has a nearby companion galaxy NGC4298 with apparently no sign of interactions in the optical images. However, radio observations with GMRT 330 MHz by Kantharia et al. (2005) shows quite a clear sign of interactions in terms of tidal bridge between the two galaxies. This galaxy has been studied recently in great detail to understand the distinct nature of extra-planar diffused ionized gas by Heald et al. (2007). We have adopted the major axis H$\footnotesize I$ rotation curve of this galaxy from Heald et al. (2007). The galaxy shows a mild warping of the underlying thin disc in the 4.5 micron band. The warp has also been seen in the optical R-band analysis (Schwartzkopf \& Dettmar 2001). The theoretically estimated onset radius using the disc (without flaring) and halo parameters overestimates the onset radius by 1.33 scale length. The measured flaring rate is very high for this galaxy. The scale height increases by a factor of $\sim$ 3 over 10 disc scale lengths. Even using the measured flaring, a discrepancy of $\sim 0.3 R_d$ remains between the theory and observations.

\clearpage

\begin{figure*}
  \setlength{\figwidth}{\textwidth}
  \addtolength{\figwidth}{-\columnsep}
  \setlength{\figwidth}{0.5\figwidth}
  
  \begin{minipage}[t]{\figwidth}
    \mbox{}
    \vskip -7pt
    \centerline{\includegraphics[width=1.0\linewidth,angle=0]{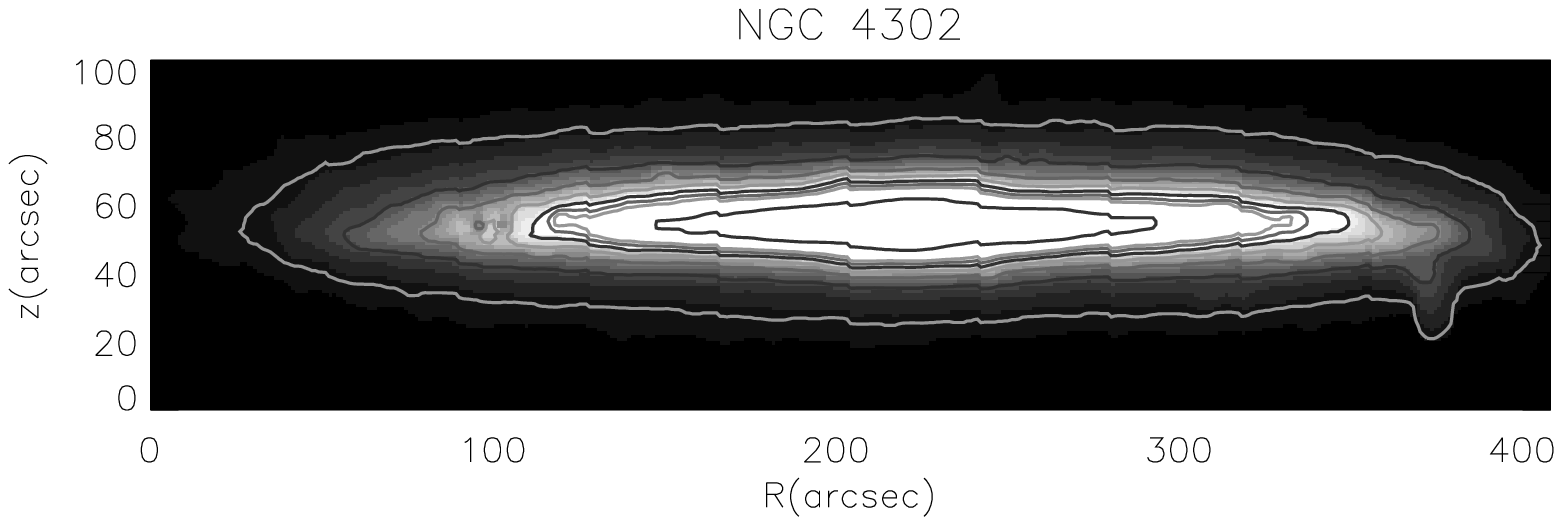}}
  \end{minipage}
  \hfill
  \begin{minipage}[t]{\figwidth}
    \mbox{}
    \vskip -7pt
    \centerline{\includegraphics[width=0.72\linewidth,angle=270]{pgvc_n4302.ps}}
  \end{minipage}
\begin{minipage}[t]{\figwidth}
    \mbox{}
    \vskip -7pt
    \centerline{\includegraphics[width=1.0\linewidth,angle=0]{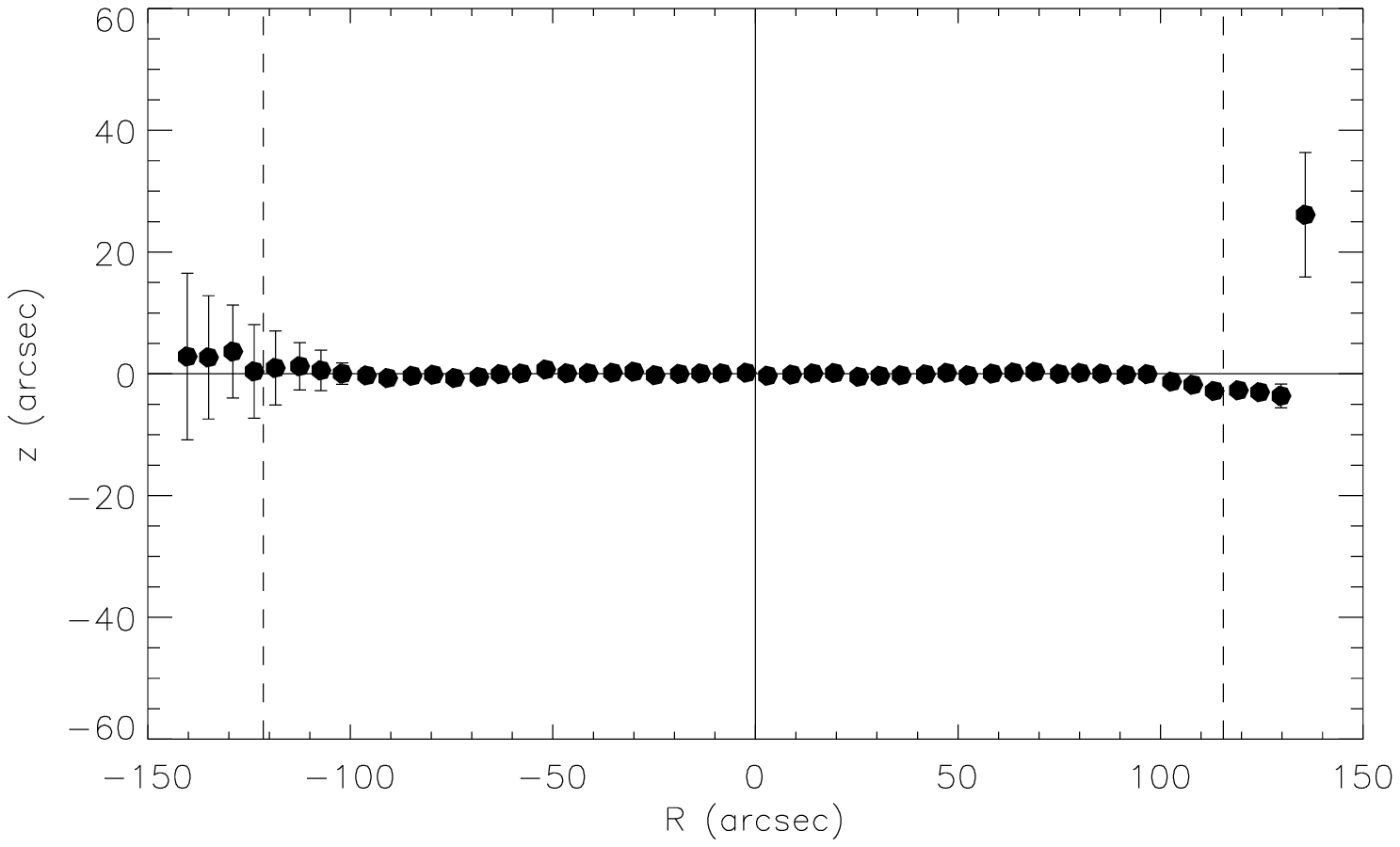}}
  \end{minipage}
  \hfill
  \begin{minipage}[t]{\figwidth}
    \mbox{}
    \vskip -7pt
    \centerline{\includegraphics[width=0.72\linewidth,angle=270]{pgresp_n4302.ps}}
  \end{minipage}
    \caption{ Same as Fig.~A1 but for NGC 4302 }

\end{figure*}


\clearpage

{\bf{NGC 4565:}}

NGC4565 is a very well studied galaxy in the literature. The disc in the 4.5 micron band is clearly warped. It also has an extended optical warp (Naeslund \& Joersaeter 1997). The thickness parameter (see Table 4) indicates that the disc is fairly thin. Using the disc parameters, obtained from the photometric analysis, we model the CO + HI rotation curve obtained from Sofue (1997). The modelled bulge of the galaxy is quite massive, but NGC4565 is a disc dominated galaxy (see Table 1). Using the disc-halo parameters derived from the modeling of the rotation curve, we calculate the theoretical radius for the onset of warp which is about 0.9 scale length more than the observed one (see Table 4).    
\noindent The stellar scale height of NGC4565 is known to be moderately flaring (Narayan \& Jog, 2002). The measured flaring rate is shown in Table 4. Using this flaring rate, we get a reasonably good match with the observation.   

\clearpage

\begin{figure*}
  \setlength{\figwidth}{\textwidth}
  \addtolength{\figwidth}{-\columnsep}
  \setlength{\figwidth}{0.5\figwidth}
  
  \begin{minipage}[t]{\figwidth}
    \mbox{}
    \vskip -7pt
    \centerline{\includegraphics[width=1.0\linewidth,angle=0]{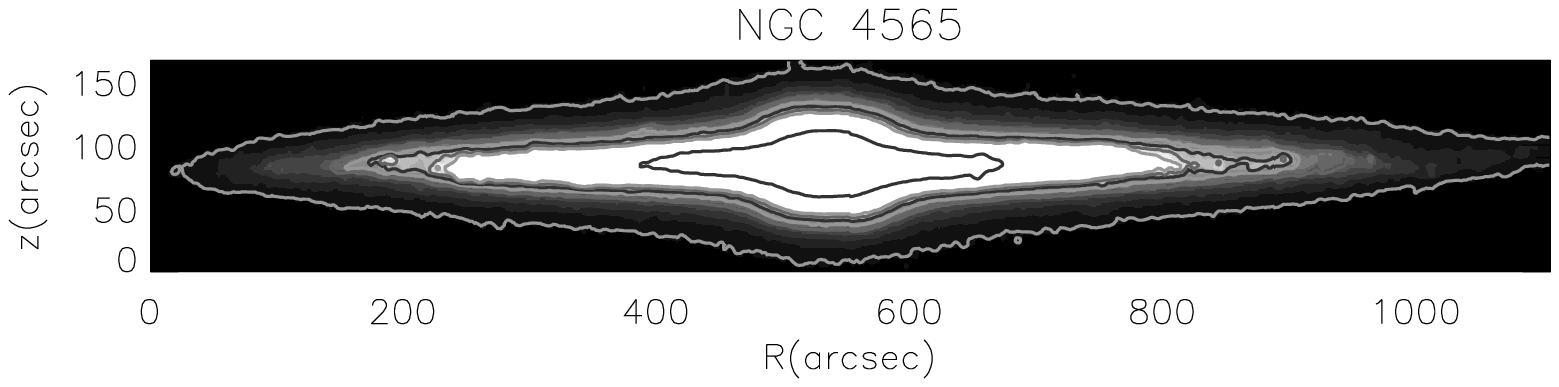}}
  \end{minipage}
  \hfill
  \begin{minipage}[t]{\figwidth}
    \mbox{}
    \vskip -7pt
    \centerline{\includegraphics[width=0.72\linewidth,angle=270]{pgvc_n4565.ps}}
  \end{minipage}
\begin{minipage}[t]{\figwidth}
    \mbox{}
    \vskip -7pt
    \centerline{\includegraphics[width=1.0\linewidth,angle=0]{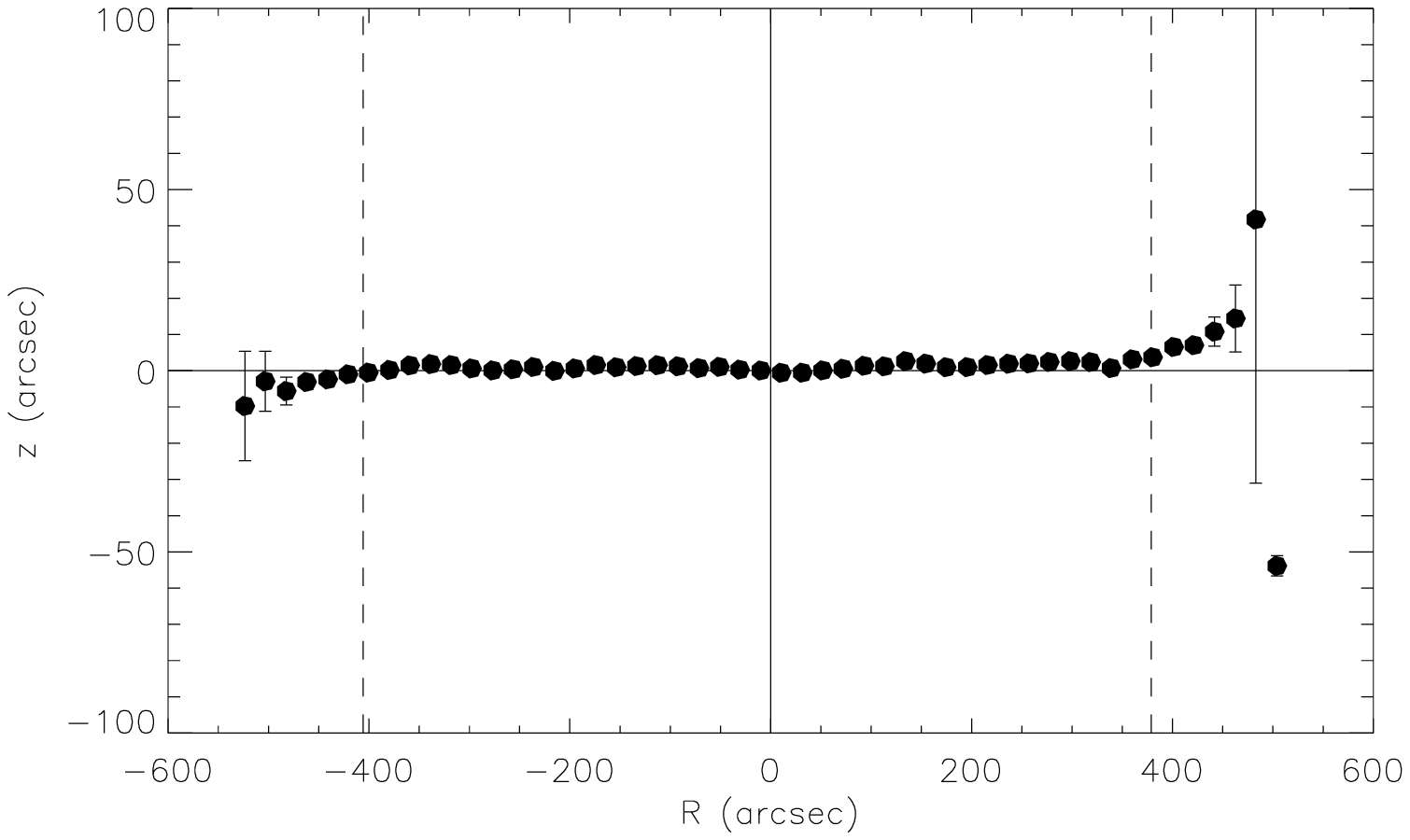}}
  \end{minipage}
  \hfill
  \begin{minipage}[t]{\figwidth}
    \mbox{}
    \vskip -7pt
    \centerline{\includegraphics[width=0.72\linewidth,angle=270]{pgresp_n4565.ps}}
  \end{minipage}
    \caption{Same as Fig.~A1 but for NGC 4565 }

\end{figure*}

\clearpage

{\bf{NGC 5907:}}

This is another well studied, nearby edge-on spiral galaxy in our sample. Warps in this galaxy have been observed both in optical $R$-band (Florido et al. 1992; Schwartzkopf \& Dettmar 2001) and in H$\footnotesize I$ (Sancisi 1976). We detect a small warp in the 4.5 micron band. Earlier thought to be a non-interacting galaxy, this galaxy has recently been discovered to be interacting with a dwarf companion PGC54419 lying at a projected distance of only $\sim$ 36 kpc from the center of the galaxy (Shang et al. 1998; Martinez-Delgado et al. 2008). The dwarf companion could to be the cause of NGC5907's warp. The disc of NGC5907 also show small scale corrugations as can be seen in the warp curve. The thickness parameter indicates that this is a fairly thin galaxy (see Table 4). 

We model the CO+HI rotation curve from Sofue (1997). The bulge of this galaxy is moderately massive. There is a large discrepancy between the theoretically estimated onset radius without any flaring and the observed one. The stellar scale height flares with a rate of 0.24 per scale length in the galaxy (see Table 4). Even using the measured flaring rate, there remains a discrepancy of about 0.5 scale length in the onset radius between the observation and theory.

\begin{figure*}
  \setlength{\figwidth}{\textwidth}
  \addtolength{\figwidth}{-\columnsep}
  \setlength{\figwidth}{0.5\figwidth}
  
  \begin{minipage}[t]{\figwidth}
    \mbox{}
    \vskip -7pt
    \centerline{\includegraphics[width=1.0\linewidth,angle=0]{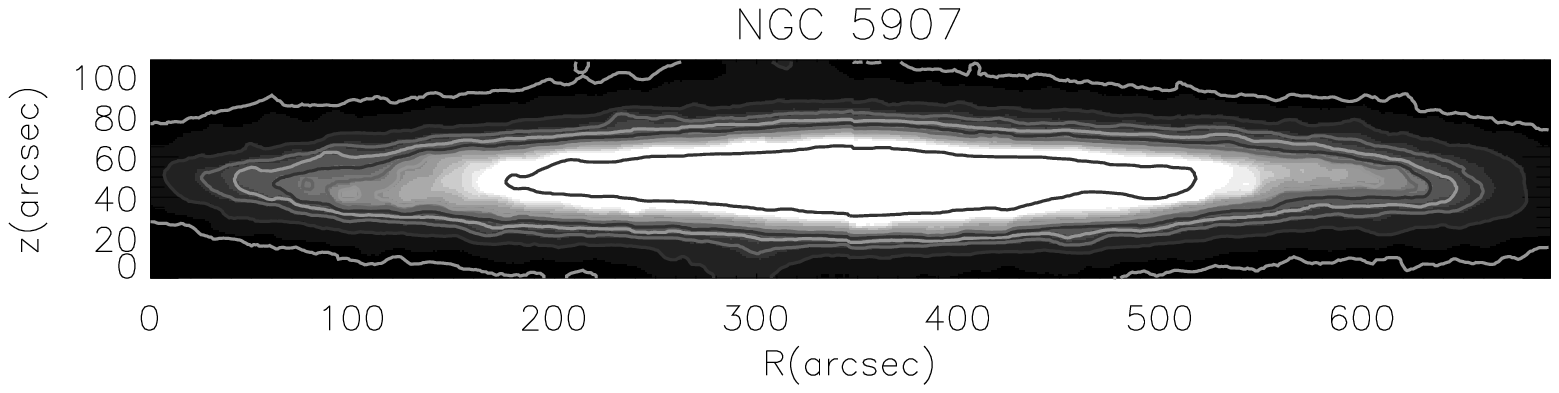}}
  \end{minipage}
  \hfill
  \begin{minipage}[t]{\figwidth}
    \mbox{}
    \vskip -7pt
    \centerline{\includegraphics[width=0.72\linewidth,angle=270]{pgvc_n5907.ps}}
  \end{minipage}
\begin{minipage}[t]{\figwidth}
    \mbox{}
    \vskip -7pt
    \centerline{\includegraphics[width=1.0\linewidth,angle=0]{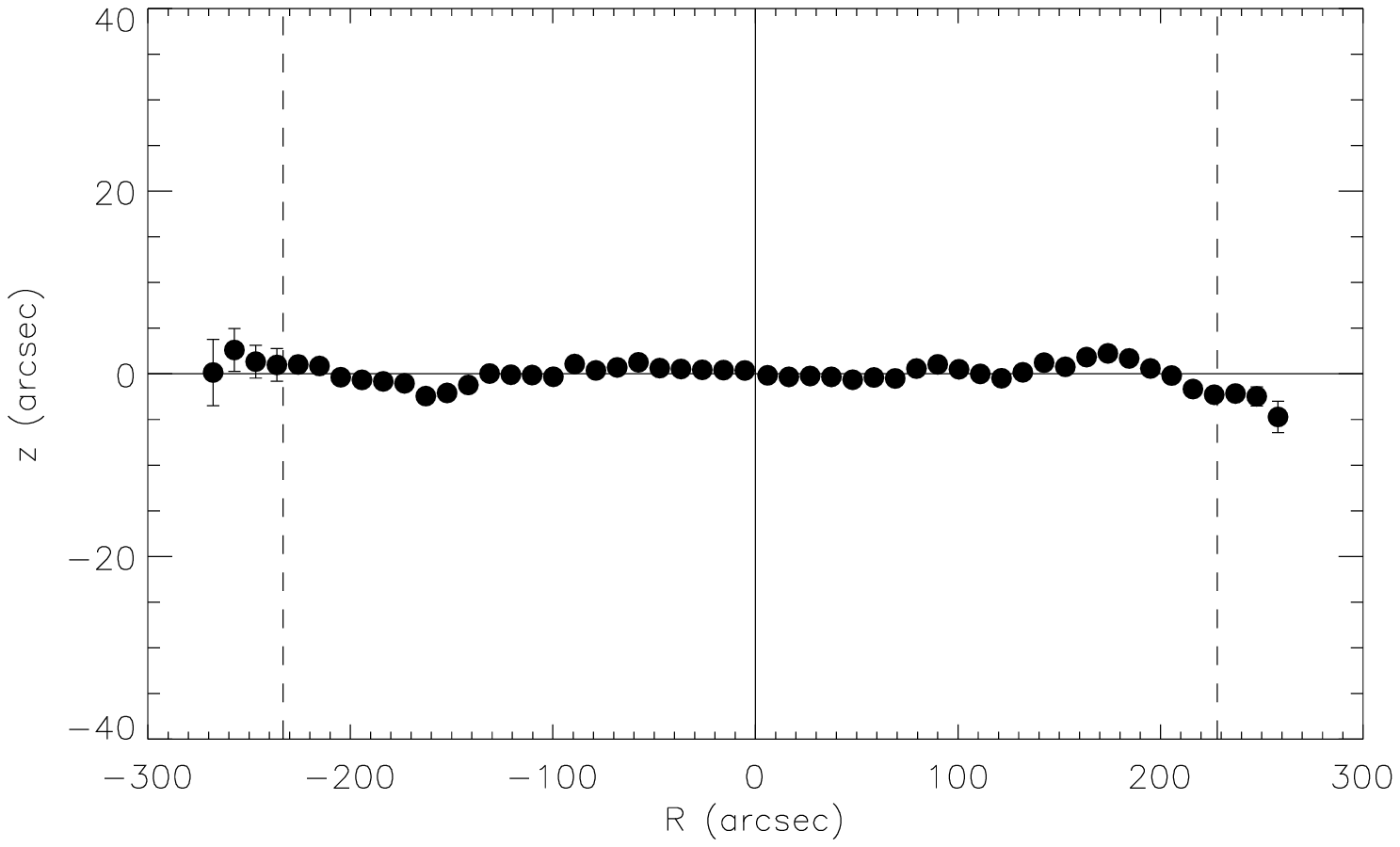}}
  \end{minipage}
  \hfill
  \begin{minipage}[t]{\figwidth}
    \mbox{}
    \vskip -7pt
    \centerline{\includegraphics[width=0.72\linewidth,angle=270]{pgresp_n5907.ps}}
  \end{minipage}
    \caption{Same as Fig.~A1 but for NGC 5907 }

\end{figure*}


\end{document}